**THERMODYNAMICS OF MIXTURES CONTAINING A FLUORINATED BENZENE AND A HYDROCARBON**

Juan Antonio González,[a*] Luis Felipe Sanz,[a] Fernando Hevia,[b] Isaías García de la Fuente,[a] and José Carlos Cobos[a]

[a]G.E.T.E.F., Departamento de Física Aplicada, Facultad de Ciencias, Universidad de Valladolid, Paseo de Belén, 7, 47011 Valladolid, Spain.

[b]Université Clermont Auvergne, CNRS. Institut de Chimie de Clermont-Ferrand. F-63000, Clermont-Ferrand, France

*corresponding author, e-mail: jagl@termo.uva.es; Fax: +34-983-423136; Tel: +34-983-423757

**Abstract**

Fluorobenzene, or 1,4-difluorobenzene or hexafluorobenzene + alkane mixtures and hexafluorobenzene + benzene, or + toluene, or + 1,4-dimethylbenzene systems have been investigated using thermodynamic properties from the literature and through the application of the DISQUAC and UNIFAC (Dortmund) models and the concentration-concentration structure factor ($S_{CC}(0)$) formalism. DISQUAC interaction parameters for the contacts F/alkane and F/aromatic have been determined. UNIFAC interaction parameters available in the literature for these contacts have been used along calculations. Both models predict double azeotropy for the $C_6F_6 + C_6H_6$ system, although in different temperature ranges. The $H_m^E$ values of the fluorobenzene, or 1,4-difluorobenzene + $n$-alkane systems are positive and are accurately described by the models using interaction parameters independent of the $n$-alkane. This means that no Patterson's effect exists in such mixtures. DISQUAC calculations allow state that such conclusion is still valid for $C_6F_6 + n$-alkane mixtures. DISQUAC provides better results than UNIFAC on $C_{pm}^E$ of solutions involving $n$-alkanes, or on $H_m^E$ of $C_6F_6$ + aromatic hydrocarbon systems. Mixtures with alkanes are characterized by interactions between like molecules, which are mainly of dispersive type, which is supported, e.g., by the negative $C_{pm}^E$ values of these systems. It is shown that structural effects can contribute largely to $H_m^E$. This is investigated in terms of the excess molar internal energy at constant volume, $U_{Vm}^E$, whose values are determined for the investigated solutions. For mixtures with a given $n$-alkane, the relative variation of $H_m^E$ and $U_{Vm}^E$ with the fluorohydrocarbons is different. $H_m^E$ values change in the sequence $C_6F_6 > 1,4\text{-}C_6H_4F_2 > C_6H_6 > C_6H_5F$, while $U_{Vm}^E$ changes in the order: $1,4\text{-}C_6H_4F_2 > C_6H_6 \approx C_6H_5F > C_6F_6$. $C_6F_6$ + aromatic hydrocarbon mixtures are characterized by interactions between unlike molecules as it is demonstrated by their negative $H_m^E$ values. The application of the $S_{CC}(0)$ formalism reveals that homocoordination is more important in $C_6F_6 + n$-alkane mixtures than in the corresponding systems with $C_6H_5F$, and that heterocoordination is dominant in the solutions of $C_6F_6$ with an aromatic hydrocarbon.

## 1. Introduction

There are two important effects which can be present in liquid mixtures formed by *n*-alkane and a molecule A of plate-like or of more or less globular shape [1-4]. They are the so-called Wilhem's and Patterson's effects. The former is encountered in systems where A is a flat molecule of the type 1,2,4-trimethylbenzene, or 1-methylnaphthalene, or 1,2,4-trichlorobenzene [4,5]. These solutions show decreasing values of molar excess enthalpies, $H_m^E$, when *n*, the number of C atoms in the *n*-alkane, is increased. Tentatively, such variation has been ascribed to the creation of a certain intramolecular order due to the flat component A hinders the rotational motion of the segments of the flexible molecules of longer *n*-alkanes. The Patterson's effect is encountered, e.g, in benzene, or toluene or cyclohexane, or CCl$_4$ + *n*-alkanes systems. Here, $H_m^E$ values increase in line with *n* more sharply than expected for mixtures including longer *n*-alkanes. This has been explained in terms of an extra endothermic contribution to $H_m^E$ which arises from the disruption of the local order existing in the longer *n*-alkanes. Issues related to order creation or order destruction have been extensively investigated by means of the Flory's theory [3,6]. Application of group contribution models is also very useful at this regard [7,8]. In fact, group contribution methods, which consider *n*-alkanes as homogeneous molecules, predict a more or less regular increment of $H_m^E$ for A + *n*-alkane systems. Thus, if $H_m^E$(experimental) > $H_m^E$(calculated) for mixtures including longer *n*-alkanes, then the Patterson's effect is present in such solutions. As an example, we provide a comparison between experimental $H_m^E$ values with the corresponding results obtained from the application of the UNIFAC (Dortmund) model [9] for benzene mixtures. At equimolar composition and 298.15 K, the experimental results are: $H_m^E$/J mol$^{-1}$ = 931 (*n* = 7) [10]; 1101 (*n* = 12) [11]; 1289 (*n* =16) [11], while the calculated values, in the same units, are: 848 (*n* = 7); 1009 (*n* = 12); 1087 (*n* =16). Therefore, if one is interested in an accurate representation of $H_m^E$ for this class of systems, interaction parameters depending on *n* must be considered (Figure S1, supplementary material). The zeroth approximation of DISQUAC [12,13] has been applied in this framework to the study of C$_6$H$_6$, or C$_7$H$_8$, or C$_6$H$_{12}$ or CCl$_4$ + *n*-alkanes systems [7,8,14], and it has been shown that the dispersive enthalpic parameter increases for larger *n* values. Similarly, the systems 1,2,4-trimethylbenzene, or chlorobenzene, or 1-chloronaphthalene, or bromobenzene + *n*-alkane have been also investigated using the same approach [5,15,16], although no numerical values for the interaction parameters have been provided. The graphical representation of the dispersive enthalpic parameter vs. *n* shows that the former decreases when the latter increases, which has been considered as a main feature of systems where the Wilhem's effect exists.

$H_m^E$ values of binary mixtures formed by one aromatic fluorinated compound (AFC) such as fluorobenzene, or 1,4-difluorobenzene, or hexafluorobenzene and one *n*-alkane increase in line with $n$ [15,17]. For $C_6F_6$ solutions, $H_m^E$ ($x_1 = 0.5$; $T = 298.15$ K)/J mol$^{-1}$ = 1119 ($n = 7$) and 1324 ($n = 14$) [15]. Thus, one of the aims of this work is to investigate if Patterson's effect is present in the mentioned systems, by comparing experimental results with those provided by DISQUAC. Along the work, DISQUAC calculations are compared with results obtained from the application of the UNIFAC (Dortmund) model using interaction parameters from the literature [9,18]. Fluorobenzene is a polar molecule (dipole moment, $\mu = 1.7$ D [19], while both 1,4-difluorobenzene and $C_6F_6$ have $\mu = 0$ D. However, the latter compound has a rather large quadrupolar moment (31.7 10$^{-40}$ C m$^2$ [20]), while this magnitude is $-21.0$ 10$^{-40}$ C m$^2$ for $C_6H_5F$ [20] and $-7.7$ 10$^{-40}$ C m$^2$ for 1,4-$C_6H_4F_2$ [21]. The study of systems with such characteristics is important for a better understanding of weak no covalent interactions, which are decisive for the prediction and control of organic solid-state structures [22]. With this purpose in mind, our study is extended to hexafluorobenzene + benzene, or + toluene, or + 1,4-dimethylbenzene (14DMBZ) mixtures. The $C_6F_6 + C_6H_6$ system deserves particular attention by its outstanding properties. It shows double azeotropy [23-25], negative $H_m^E$ values [26,27] ($-496$ J mol$^{-1}$ at 298.15 K and equimolar composition [27]) and, although both compounds are liquids at room temperature, form a complex of the 1:1 type in solid state [28-30], which is a typical organic crystal. Initially, the existence of this adduct was ascribed to the formation of a bond of the π–π* type between $C_6H_6$ and $C_6F_6$ [31,32]. However, spectroscopic studies do not support that the $C_6F_6$–benzene interactions are due to a charge transfer [33]. Investigations on the charge distributions in these complexes show that their cohesion arises from to interactions between their quadrupolar moments [20,34,35]. On the other hand, the mentioned thermodynamic data suggest that the complex still exists in liquid state, which is supported by detailed studies using neutron or x-diffraction techniques [36]. The selected systems for this study are also investigated in the framework of the concentration-concentration structure factor ($S_{CC}(0)$) formalism [37,38], whose purpose is to try to link the thermodynamic properties of a given liquid mixture with local deviations from the bulk composition. Particularly, the method, based on the Bhatia-Thorton partial structure factors [39], is concerned with the study of fluctuations in the number of molecules in a binary mixture regardless of the components, the fluctuations in the mole fraction and the cross fluctuations.

Aromatic fluorinated compounds have many applications. For example, fluorobenzene is used in the manufacturing of Li batteries [40]. On the other hand, aromatic compounds are commonly encountered in urban air arising from the fossil fuel combustion emissions [41] and the understanding of the atmospheric chemistry of these compounds is required to assess their environmental impact [42]. In this context, fluorobenzene is used in the development of new lower boiling fuels which protect the environment by reducing emissions. The design of new organic co-crystals is relevant in the pharmaceutical industry since can replace salts. In addition, complex aromatic fluorinated molecules are used as antidepressants, or as drugs to reduce cholesterol [43].

The study of aromatic-aromatic interactions involving fluorine substituted benzenes is also interesting in order to investigate the binding of drugs at their sites of action [19,44].

## 2. THEORIES

### 2.1 DISQUAC

DISQUAC is based on the rigid lattice theory developed by Guggenheim [45]. Some relevant features of the model are now described. (i) The total molecular volumes, $r_i$, surfaces, $q_i$, and the molecular surface fractions, $\alpha_i$, of the compounds present in the mixture are calculated additively on the basis of the group volumes $R_G$ and surfaces $Q_G$ recommended by Bondi [46]. As volume and surface units, the volume $R_{CH4}$ and surface $Q_{CH4}$ of methane are taken arbitrarily [13]. The geometrical parameters for the groups referred to in this work are given elsewhere [13,47,48]. (ii) The partition function is factorized into two terms, and the excess functions are calculated as the sum of two contributions: a dispersive (DIS) term which represents the contribution from the dispersive forces; and a quasichemical (QUAC) term which arises from the anisotropy of the field forces created by the solution molecules. For the molar excess Gibbs energy, $G_m^E$, a combinatorial term, $G_m^{E,COMB}$, described by the Flory-Huggins equation [13] must be included. Thus,

$$G_m^E = G_m^{E,COMB} + G_m^{E,DIS} + G_m^{E,QUAC} \qquad (1)$$

$$H_m^E = H_m^{E,DIS} + H_m^{E,QUAC} \qquad (2)$$

(iii) The interaction parameters are assumed to be dependent on the molecular structure. (iv) The coordination number $z = 4$ is used for all the polar contacts. This is an important shortcoming of the model, and is partially removed via the hypothesis of considering structure dependent interaction parameters. (v) It is also assumed that there is no volume change upon mixing, that is, $V_m^E$, the excess molar volume, is 0.

The equations used to calculate the DIS and QUAC contributions to $G_m^E$ and $H_m^E$ in the framework of DISQUAC are given elsewhere [12,49]. The temperature dependence of the interaction parameters is expressed in terms of the DIS and QUAC interchange coefficients [12, 49], $C_{st,l}^{DIS}; C_{st,l}^{QUAC}$ where s ≠ t are two contact surfaces present in the mixture and $l = 1$ (Gibbs energy; $C_{st,1}^{DIS/QUAC} = g_{st}^{DIS/QUAC}(T_o)/RT_o$); $l = 2$ (enthalpy, $C_{st,2}^{DIS/QUAC} = h_{st}^{DIS/QUAC}(T_o)/RT_o$)); $l = 3$ (heat capacity, $C_{st,3}^{DIS/QUAC} = c_{pst}^{DIS/QUAC}(T_o)/R$)). $T_o$ = 298.15 K is the scaling temperature and $R$, the gas constant.

*2.2 UNIFAC (Dortmund)*

Modified UNIFAC (Dortmund version) differs from original UNIFAC [50] by the combinatorial term and the temperature dependence of the interaction parameters. The equations needed to calculate $G_m^E$ and $H_m^E$ are obtained from the fundamental equation for the activity coefficient $\gamma_i$ of component i:

$$\ln \gamma_i = \ln \gamma_i^{COMB} + \ln \gamma_i^{RES} \tag{3}$$

where $\ln \gamma_i^{COMB}$ and $\ln \gamma_i^{RES}$ are, respectively, the combinatorial and residual terms. Equations can be found elsewhere [49]. On the other hand, in this UNIFAC version, the geometrical parameters, the relative van der Waals volumes, and the relative van der Waals surfaces of the different groups/subgroups are not calculated from molecular parameters like in the original UNIFAC, but fit together with the interaction parameters to the experimental values of the thermodynamic properties considered. In the framework of UNIFAC (Dortmund), AFCs are represented by the main group ACF (Nº 38). No subgroups are defined within this main group in such way that, e.g., hexafluorobenzene is considered as an homogeneous molecule. The geometrical and interaction parameters were taken from the literature and used without modification [9,18].

*2.3 The concentration-concentration structure factor formalism*

For a binary system, the $S_{CC}(0)$ function can be determined from the equation [37,38]:

$$S_{CC}(0) = \frac{x_1 x_2}{1 + \frac{x_1 x_2}{RT}\left(\frac{\partial^2 G_m^E}{\partial x_1^2}\right)_{P,T}} \tag{4}$$

In the case of ideal mixtures, $G_m^{E,id} = 0$; and $S_{CC}^{id}(0) = x_1 x_2$. Stability conditions require that $S_{CC}(0) > 0$. Thus, if a system is close to phase separation, $S_{CC}(0)$ must be large and positive and the dominant trend is the separation between components (homocoordination), and $S_{CC}(0) > x_1 x_2$. If compound formation between components exists (heterocoordination), $S_{CC}(0)$ must be very low and $0 < S_{CC}(0) < x_1 x_2$. More details can be encountered in reference [37].

3. **Estimation of DISQUAC interaction parameters**

In the framework of DISQUAC, $C_6H_5F$, or 1,4-$C_6H_4F_2$, or $C_6F_6$ + *n*-alkane, or + cyclohexane, or + methylcyclohexane, or + benzene, or + toluene, or + 14DMBZ mixtures are regarded as possessing the following four types of surfaces: (i) type a, aliphatic ($CH_3$, $CH_2$, in *n*-alkanes, methylcyclohexane, toluene or 14DMBZ); (ii) type b, aromatic ($C_6H_6$ in benzene, $C_6H_5$- in toluene or fluorobenzene, $C_6H_4$- in 14DMBZ or 1,4-difluorobenzene, or aromatic ring in $C_6F_6$); (iii) type c, c-$CH_2$ or c-CH in cyclohexane or methylcyclohexane; (iv) type f, F in $C_6H_5F$, 1,4-$C_6H_4F_2$ or $C_6F_6$. Detailed explanations on the general procedure applied in the estimation of the DISQUAC interaction parameters have been given elsewhere [12,49]. Some important remarks are provided below.

*3.1 The $C_6H_5F$, or 1,4-$C_6H_4F_2$, or $C_6F_6$ + $C_6H_6$ systems*

These mixtures are characterized only by the (b,f) contact. In view of the thermodynamic properties of the solution containing $C_6F_6$, described in Introduction, we have assumed that the mentioned contact in this solution is represented by DIS and QUAC interaction parameters (Table 1). The $C_6H_5F$, or 1,4-$C_6H_4F_2$ + $C_6H_6$ systems show low $H_m^E$ and $G_m^E$ values. For example, at equimolar composition and 303.15 K, $H_m^E$/J mol$^{-1}$ = 1 ($C_6H_5F$) and 65 (1,4-$C_6H_4F_2$) [51]. In addition, for the $C_6H_5F$ + $C_6H_6$ system at $x_1$ = 0.5 and 348.15 K, $G_m^E$ = 27 J mol$^{-1}$ [52]. These experimental results suggest that the corresponding (b,f) contacts can be represented by DIS parameters only. Calculations showed that the $C_{bf,l}^{DIS}$ (l =1,2,3) coefficients are negligible. Thus, for the sake of simplicity, we have assumed that $C_{bf,l}^{DIS}$ = 0 (l =1,2,3) for the mixtures with $C_6H_5F$, or 1,4-$C_6H_4F_2$.

*3.2 The $C_6H_5F$, or 1,4-$C_6H_4F_2$, or $C_6F_6$ + alkane systems*

Three types of contacts exist in these mixtures: (s,b), (s,f) and (b,f), where s = a, aliphatic in *n*-alkanes or methylcyclohexane; or s = c, c-$CH_2$ or c-CH in cyclohexane or methylcyclohexane. The (s,b) contacts are described by dispersive interaction parameters only and are known from the study of alkylbenzene + alkane mixtures for systems with toluene, or 14DMBZ [47]. However, interaction parameters are not available for solutions including hexamethylbenzene and a survey of literature data using the ThermoLit application showed that there is a lack of the required experimental data to conduct a fitting of the interaction parameters for $C_6(CH_3)_6$ + alkane mixtures. As a consequence, we have considered that $C_{sb,l}^{DIS}$ = 0 (s = a,c; l =1,2,3) for systems with $C_6F_6$. Regarding the (b,f) contacts, the interaction parameters are already known (see above). The parameters for the (a,f) contacts (Table 1) were then determined under the following conditions: (i) the contacts are represented by DIS and QUAC interaction parameters. (ii) The QUAC parameters are considered to be the same for mixtures with $C_6H_5F$ or 1,4-$C_6H_4F_2$. The adjustment of the (c,f) parameters was conducted assuming that that $C_{af,l}^{QUAC} = C_{cf,l}^{QUAC}$ (l =1,2,3). In the framework of DISQUAC, this is a typical procedure applied when investigating, e.g., mixtures formed by ether [53] or chloroalkane [54], or *n*-alkanone [55], or linear organic carbonate [56] or tertiary amide [57], or 1-alkanol [58] and *n*-alkane or cyclohexane.

*3.3 The $C_6F_6$ + toluene, or +1,4-dimethylbenzene systems*

These mixtures are also built by three contacts: (a,b); (a,f) and (b,f). The interaction parameters for (a,b) and (a,f) contacts are already known. The determination of the parameters for the (b,f) contacts was performed under the basic assumption that the systems with benzene or toluene or 1,4-dimethylbenzene are characterized by the same QUAC coefficients but different DIS parameters (see Table 1). A similar procedure we applied in previous studies concerned with mixtures formed by tertiary amide [57], or 1-alkanol [59], or 2-alkanol [60], or sulfolane [61] and an aromatic hydrocarbon.

## 4. Theoretical results

Results from DISQUAC on phase equilibria, $H_m^E$ and $C_{pm}^E$ are shown in Tables 2-5 and in Figures 1-6. Tables 2 and 4 contain relative deviations for pressure and for $H_m^E$, respectively, defined as

$$\sigma_r(P) = \{\frac{1}{N}\sum\left[\frac{P_{exp}-P_{calc}}{P_{exp}}\right]^2\}^{1/2} \qquad (5)$$

$$dev(H_m^E) = \{\frac{1}{N}\sum\left[\frac{H_{m,exp}^E - H_{m,calc}^E}{H_{m,exp}^E(x_1=0.5)}\right]^2\}^{1/2} \qquad (6)$$

where $N$ stands for the number of data points. Similarly, results from the application of the UNIFAC model are listed in Tables 2,4 and 5.

Table 6 contains DISQUAC calculations on $S_{CC}(0)$ (see Figure 7), and a comparison between theoretical results with experimental values determined in this work using VLE data from the literature (see also Figure S2 of supplementary material). We note that DISQUAC is a reliable tool to evaluate the $S_{CC}(0)$ magnitude.

Results on the double azeotropy of the $C_6F_6 + C_6H_6$ system deserve some comments. Experimental data show that two azeotropes are encountered in the temperature range (303.15-453.15) K [23-25]. UNIFAC describes correctly this behaviour between (303.15-363.15) K. At 373.15, the model predicts only a positive azeotrope at $x_{1az} = 0.25$ and $P_{az} = 188.2$ kPa. DISQUAC only predicts the existence of positive azeotropes at the lower temperatures, i.e. in the range (303.15-350.15) K. From 353.45 K to 413.15 K, DISQUAC correctly predicts the existence of two azeotropes. In fact, at 413.15 K, we obtain $x_{1az}$ = 0.25, $P_{az}$ = 505.1 kPa and $x_{1az}$ = 0.60, $P_{az}$ = 501 kPa. The $C_6F_6$ + $C_6H_{12}$ system has a eutectic point at $x_{1eu}$ = 0.185 and $T_{eu}$ = 251.6 K [28]. The DISQUAC results, $x_{1eu}$ = 0.162 and $T_{eu}$ = 252.3 K, show the model can be applied in a wide range of temperature. On the other hand, neglecting ternary interactions, i.e., using binary parameters only, DISQUAC provides the following $\sigma_r(P)$ values for the VLE at 298.15 K of the ternary systems $C_6F_6$ + $C_6H_6$ + tetradecane ($N$ = 27) [62] or + hexadecane ($N$ = 29) [63]: 0.040 and 0.045, respectively, which are very close to UNIFAC results (0.049 and 0.041).

## 5. Discussion

Below, we are referring to values of the thermodynamic properties at equimolar composition and 298.15 K.

*5.1 Alkane mixtures*

These systems show positive $H_m^E$ values (Table 4, see below), and therefore are characterized by interactions between like molecules. As already mentioned, $\mu$, of $C_6H_5F$ is 1.7 D [19], while $C_6F_6$ and 1.4-$C_6H_4F_2$ have $\mu$ = 0 D [19]. However, ($C_{pm}^E$ / J mol$^{-1}$ K$^{-1}$) values of $C_6H_5F$ or $C_6F_6$ + alkane mixtures are negative ($-1.12$ [17] and $-1.18$ [15], respectively, for the corresponding heptane systems, see Table 5), which indicates that dispersive interactions are dominant. Negative $C_{pm}^E$ values are typically ascribed to order destruction [1-3]. This is also the case of $C_6H_6$, or $C_6H_{12}$ + $n$-alkane systems. For example, $C_{pm}^E$/J mol$^{-1}$ K$^{-1}$ = $-3.34$ for the benzene + heptane mixture [64]. The importance of dispersive interactions in the investigated solutions is supported by their situation in the $G_m^E$ vs. $H_m^E$ diagram [65-67]. In fact, they are placed between the lines $G_m^E$ = 1/3 $H_m^E$ and $G_m^E$ = 1/2 $H_m^E$, or very close to this region, which is where mixtures characterized by dispersive interactions are encountered. We give some $G_m^E$ and $H_m^E$ values for reference. Thus, for heptane systems: $G_m^E$/J mol$^{-1}$ = 430 ($C_6H_6$) [68]; 435 ($C_6H_5F$); 488 (1,4-$C_6H_4F_2$), 644 ($C_6F_6$) (DISQUAC values) and $H_m^E$/J mol$^{-1}$ = 931 ($C_6H_6$) [10]; 888 ($C_6H_5F$) [17]; 1050 (1,4-$C_6H_4F_2$) [15] and 1119 ($C_6F_6$) [15]. Obviously, all the systems show positive $TS_m^E (= H_m^E - G_m^E)$ values, which is also a main feature of systems where dispersive interactions are dominant. In addition, it should be mentioned that a number of $H_m^E$ curves are skewed towards high concentration of the AFC (see Figures 2 and 3 and S1 of supplementary material). This is important since it seems to be a manifestation of the existence of a quadrupolar order [2,69]. In the case of polar + non-polar systems, $H_m^E$ curves are predicted to be shifted to lower concentrations of the polar compound [2,69]. Finally, preliminary calculations on the basis of the Flory model reveal that orientational effects in these solutions are weak. In fact, we have obtained: $dev(H_m^E)$ (heptane) = 0.021 ($C_6H_5F$); 0.082 (1,4-$C_6H_4F_2$) and 0.074 ($C_6F_6$) and $dev(H_m^E)$ (hexadecane) = 0.016 ($C_6H_5F$); 0.020 (1,4-$C_6H_4F_2$) and $dev(H_m^E) = 0.017$ for the $C_6F_6$ + tetradecane mixture.

*5.2 Mixtures with a given fluorobenzene*

In this case, $H_m^E$ increases with $n$ (Table 4, Figures 2-4). The comparison between DISQUAC values for mixtures including $C_6H_5F$ or 1,4-$C_6H_4F_2$ with experimental data show an excellent agreement between them (Table 4, Figures 2-3). Therefore, one can conclude that there is no Patterson's effect, since calculations were performed using enthalpic parameters independent of the *n*-alkane. This is confirmed by the results obtained from UNIFAC (Dortmund) calculations (Table 4). A similar conclusion can be stated for $C_6F_6$ mixtures on the basis of the results provided by DISQUAC (Table 4, Figure 4).

For the present solutions, $C_{pm}^E$ becomes progressively more negative when $n$ is increased (Table 5, Figures 5 and 6). DISQUAC describes accurately such behaviour by means of $C_{af,3}^{DIS}$ coefficients which depend on $n$ (Table 1). This merely expresses that $C_{pm}^E$ is a property closely related to the mixture structure. Nevertheless, we remark that the model still predicts the $C_{pm}^E$ variation with $n$ using a unique $C_{af,3}^{DIS}$ value. In the case of $C_6H_5F$ systems, if $C_{af,3}^{DIS} = -0.5$, then DISQUAC yields: $C_{pm}^E$/J mol$^{-1}$ K$^{-1}$ = $-0.9$ ($n = 7$); $-1.4$ ($n = 8$); $-2.0$ ($n = 10$); $-3.1$ ($n = 12$); $-4.4$ ($n = 14$), in fair agreement with experimental results (Table 5).

$V_m^E$ data also increase with $n$ [15,17] (Table 7). For example, $V_m^E$ ($C_6H_5F$)/cm$^3$ mol$^{-1}$ = 0.246 ($n = 6$); 0.833 ($n = 16$) [17]. That is, both $H_m^E$ and $V_m^E$ functions have the same sign and increase in line (Table 7). This underlines the importance of the interactional contribution to $V_m^E$, which is supported by the fact that the $H_m^E$ and $V_m^E$ curves show a similar symmetry. From the few data available on $V_m^E(T)$, we obtain $\frac{\Delta V_m^E}{\Delta T}$/ cm$^3$ mol$^{-1}$ K$^{-1}$ = 0.0016 ($C_6H_5F$ + methylcyclohexane) [70,71]; 0.0017 ($C_6H_5F$ + $C_6H_{12}$) [71,72] and 0.008 ($C_6F_6$ + $C_6H_{12}$) [73,74]. For the $C_6H_6$ + $C_6H_{12}$ system, this value is 0.00064 cm$^3$mol$^{-1}$ K$^{-1}$.[71,75]. The mixtures with $C_6H_5F$ or $C_6F_6$ are more sensitive to temperature changes probably because these liquids are more ordered.

It is to be noted that the $C_6H_5F$ + hexane, or + $C_6H_{12}$ systems show the same $H_m^E$ results (869 [17] and 864 [71] J mol$^{-1}$, respectively) but very different ($V_m^E$/cm$^3$ mol$^{-1}$) values (0.246 [17] and 0.686 [70], in the same order as above), which underlines the importance of shape effects on $V_m^E$. We remark that some mixtures involving 1,4-$C_6H_4F_2$ or $C_6F_6$ show very large and positive $V_m^E$ values (Table 7). For example, $V_\text{m}^\text{E}$ ($C_6F_6$)/cm$^3$ mol$^{-1}$= 1.882 ($n$ = 7); 2.093 ($n$ = 14) [15]; 2.454 (cyclohexane [73]). In such cases, one can expect that the equation of state contribution (eos) to $H_\text{m}^\text{E}$ is also large (see below). In the following, we'll see that it is crucial to take into account this fact in order to conduct a consistent discussion.

*5.3 Systems with a given alkane*

At this condition, $H_\text{m}^\text{E}$/J mol$^{-1}$ of mixtures with heptane changes in the sequence: 888 ($C_6H_5F$) [17] < 1050 (1,4-$C_6H_4F_2$) [15] < 1119 ($C_6F_6$) [15]. The partial excess molar enthalpies at infinite dilution of the first compound (fluorinated benzene), $H_{\text{m},1}^{\text{E},\infty}$/kJ mol$^{-1}$ determined from $H_\text{m}^\text{E}$ measurements for the same systems over the entire composition range are: 3.48 [17]; 3.47 and 4.31 [15], respectively. However, these results do not provide a reliable information about interactions between AFC molecules due to the mentioned large eos contribution to $H_\text{m}^\text{E}$. For this reason, we conduct now a discussion in terms on the excess molar internal energies at constant volume, $U_{V\text{m}}^\text{E}$. This magnitude can be determined from [76]:

$$U_{V\text{m}}^\text{E} = H_\text{m}^\text{E} - T\frac{\alpha_p}{\kappa_T}V_\text{m}^\text{E} \qquad (7)$$

where $\alpha_p$ and $\kappa_T$ are, respectively, the isobaric thermal expansion coefficient and the coefficient of isothermal compressibility of the system under consideration. The $T\frac{\alpha_p}{\kappa_T}V_\text{m}^\text{E}$ term is defined as the contribution from the eos term to $H_\text{m}^\text{E}$. The $\alpha_p$ and $\kappa_T$ values have been calculated here assuming ideal behavior for the mixtures ($M^\text{id} = \phi_1 M_1 + \phi_2 M_2$; with $M_i = \alpha_{pi}$, or $\kappa_{T_i}$ and $\phi_i = x_i V_{\text{m},i}/(x_1 V_{\text{m},1} + x_2 V_{\text{m},2})$). For alkanes, their $\alpha_{pi}$, and $\kappa_{T_i}$ values were taken from the literature [77]. For aromatic compounds, these values are listed in Table S1 (supplementary material). Table 7 contains our results on $U_{V\text{m}}^\text{E}$. We note that the eos contribution to $H_\text{m}^\text{E}$ represents ca. 50% in the case of systems with $C_6F_6$. For heptane mixtures, $U_{V\text{m}}^\text{E}$ changes as follows: 557 ($C_6F_6$) < 765 ($C_6H_5F$) < 854 (1,4-$C_6H_4F_2$) (all values in J mol$^{-1}$). In addition, $U_{V\text{m},1}^{\text{E},\infty}$ / kJ mol$^{-1}$ = 2.94 ($C_6H_5F$); 2.65 (1,4-$C_6H_4F_2$) and 2.42 ($C_6F_6$). The corresponding value for the $C_6H_6$ + heptane mixture is 2.7 kJ mol$^{-1}$ (value obtained using $H_\text{m}^\text{E}$ and $V_\text{m}^\text{E}$ data from references [10] and [78], respectively). From these results, some conclusions can be stated. (i) Although interactions between fluorohydrocarbon molecules are quite similar, they are stronger between $C_6H_5F$ molecules. (ii) The fact that $U_{V\text{m}}^\text{E}$(1,4-$C_6H_4F_2$) > $U_{V\text{m}}^\text{E}$($C_6H_5F$) should be interpreted as a consequence of that more dispersive interactions are broken upon mixing in 1,4-$C_6H_4F_2$ systems. (iii) Although $U_{Vm,1}^{E,\infty}$ is lower for $C_6F_6$ solutions, the lowest $U_{Vm}^E$ values obtained for these mixtures suggest that *n*-alkanes are not good breakers of the $C_6F_6$ network. At this regards, the $C_6F_6$ + 2,2,4-trimethylpentane (TMP) mixture is very interesting. This solution shows a large and positive $H_m^E$ value (1740 J mol$^{-1}$) although it is affected by rather large error (ca. 20%) since is determined from $G_m^E$ data at different temperatures [79]. The mentioned result is higher than that for the $C_6F_6$ + heptane mixture (see above). The $U_{Vm}^E$ value of the $C_6F_6$ + TMP is also higher (1355 J mol$^{-1}$, value determined using $V_m^E$ = 1.375 cm$^3$ mol$^{-1}$ [79]). One can conclude that TMP is a good breaker of the existing order in $C_6F_6$. It may be pertinent to compare these results with those for $C_6H_6$ mixtures. For such systems, the eos contribution to $H_m^E$ is also large, particularly for solutions including longer *n*-alkanes (Table 7). The $H_m^E$ variation follows the order: $C_6F_6$ > 1,4-$C_6H_4F_2$ > $C_6H_6$ > $C_6H_5F$, while $U_{Vm}^E$ changes in the order: 1,4-$C_6H_4F_2$ > $C_6H_6$ ≈ $C_6H_5F$ > $C_6F_6$. It is to be noted, that $C_6H_6$ and $C_6H_5F$ systems behave quite similarly. On the other hand, it is remarkable that $U_{Vm}^E$ varies with *n* more smoothly for systems including $C_6F_6$ than for mixtures with, e.g. $C_6H_5F$. The linear regressions of $U_{Vm}^E$ data vs. *n* give slopes, which are (in J mol$^{-1}$) 11.9 and 19.2, respectively. This seems to indicate that *n*-alkanes are not good breakers of the $C_6F_6$ network.

Mixtures with $C_6H_{12}$ behave differently and both $H_m^E$ and $U_{Vm}^E$ values of $C_6H_5F$ systems are lower than for systems containing $C_6F_6$. We have already mentioned that $H_m^E$ values of $C_6H_5F$ + hexane, or + $C_6H_{12}$ mixtures are very similar. However, $H_m^E$ is much higher for the $C_6F_6$ + $C_6H_{12}$ mixture (1576 J mol$^{-1}$ [26]) than for the solution with heptane, and $U_{V,m}^E$ is also higher for the former solution (Table 7). This result contrasts with that provided above for the TMP mixture. Both cyclohexane and TMP are globular molecules and one could expect similar results for the systems with $C_6F_6$. It seems clear that the ability of TMP for destructing the $C_6F_6$ network is much higher

*5.4 $C_6F_6$ + aromatic hydrocarbon systems*

At 313.2 K, the $H_m^E$/J mol$^{-1}$ values of these mixtures are negative (−433 ($C_6H_6$); −1140 ($C_7H_8$); −1660 (1,4-$C_6H_4(CH_3)_2$ [27], Figure 4). This means that interactions between unlike molecules are dominant. The following features support this statement. (i) The solutions are placed in the third quarter of the the $G_m^E$ vs. $H_m^E$ diagram, since both magnitudes are negative (Tables 2 and 4). The 1-alkanol + linear primary amine systems, characterized by strong interactions between unlike molecules [49], are situated in the same region. (ii) From $H_m^E$ measurements at different temperatures, it is obtained that $C_{pm}^E$ is positive, which is indicative of order creation [1-3]. This is particularly relevant for the systems with C$_7$H$_8$ or 14DMBZ, whose values are 16.7 and 12.3 J mol$^{-1}$ K$^{-1}$, respectively [27]. (iii) Accordingly, $TS_m^E$ values are negative.

At 313.15 K, we note that $V_m^E$/cm$^3$ mol$^{-1}$ decreases when the number of the CH$_3$ groups attached to the aromatic ring increases: 0.801 (C$_6$H$_6$), 0.416 (C$_7$H$_8$), 0.0855 (14DMBZ) [74]. That is, $H_m^E$ and $V_m^E$ change in line, which underlines the relevance of the interactional contribution to $V_m^E$. Since structural effects contribute negatively to $V_m^E$, the positive values of this magnitude may indicate that the positive contribution to $V_m^E$ from the disruption of interactions between like molecules largely exceeds to those emerging from both structural effects and from the creation of interactions between unlike molecules.

It is noteworthy that the $H_m^E$ curve of the C$_6$F$_6$ + C$_6$H$_6$ mixture shows a small region at low mole fractions of C$_6$F$_6$ with positive $H_m^E$ values (Figure 4). Such region vanishes when $U_{Vm}^E$ data are considered. Therefore, it may be ascribed to dominant structural effects in such region. The $U_{Vm}^E$ value of the benzene mixture is $-719$ J mol$^{-1}$, a very different result to the theoretical one obtained from molecular simulations ($-2700$ J mol$^{-1}$) [80].

On the other hand, our calculations show that $U_{Vm}^E$ and $H_m^E$ change in the same sequence: C$_6$H$_6$ > C$_7$H$_8$ > 14DMBZ (Table 7). Thus, interactions between unlike molecules are expected to be stronger in the 14DMBZ solution. In addition, the melting points of the 1:1 complexes formed in solid phase changes in the order: 300.6 K (14DMBZ) > 297.2 K (C$_6$H$_6$) > 283.15 K (C$_7$H$_8$) [28] and this also supports that interactions between unlike molecules are stronger in the mixture containing 14DMBZ. The lower melting point for the toluene system may be related, in some extent, to the much lower melting point of this hydrocarbon.

*5.5 The concentration-concentration structure factor formalism*

$S_{CC}(0)$ of C$_6$H$_6$ + $n$-alkane mixtures decreases when $n$ increases (Table 6, Figure S2, supplementary material). That is, interactions between like molecules are more easily disrupted when the system contains long chain $n$-alkanes. It is remarkable that $S_{CC}(0)$ values of mixtures formed by C$_6$F$_6$ and tetradecane or hexadecane are very similar and the corresponding curves are skewed towards higher C$_6$F$_6$ concentrations. These features suggest that longer $n$-alkanes are not good breakers of the C$_6$F$_6$-C$_6$F$_6$ interactions. The lower $S_{CC}(0)$ value of the C$_6$F$_6$ + TMP mixture reveals that the globular TMP molecule can disrupt more easily the C$_6$F$_6$ network. On the other hand, homocoordination is more important in mixtures with C$_6$F$_6$ than in systems with benzene. Since C$_6$F$_6$-C$_6$F$_6$ and C$_6$H$_6$-C$_6$H$_6$ interactions are characterized by similar $U_{Vm,1}^{E,\infty}$ values, the observed behaviour might be ascribed to C$_6$F$_6$ is more ordered liquid. Next, we discuss some results on $S_{CC}(0)$ from DISQUAC calculations. Firstly, the model correctly predicts the symmetry of the $S_{CC}(0)$ curves of the $C_6F_6$ + tetradecane, or + hexadecane mixtures (Figure 7). DISQUAC provides rather similar maxima values for these systems and for the heptane solution (Table 6). It is to be noted that the curve of this mixture is more symmetrical (Figure 7). This supports our previous conclusion about longer alkanes are poorer breakers of the $C_6F_6$ order. DISQUAC provides $S_{CC}(0)$ values for the $C_6H_5F$ mixtures that slightly differ from those of the $C_6H_6$ systems, which suggests that dipolar interactions are not relevant in solutions with fluorobenzene. In addition, DISQUAC predicts a higher homocoordination in $C_6F_6$ solutions, indicating that this is a more structured liquid. In the case of systems with cyclohexane, homocoordination is also higher in mixtures with $C_6F_6$, and is more relevant in mixtures with cyclohexane than in solutions containing an *n*-alkane of similar size. Finally, the $C_6F_6$ + aromatic hydrocarbon systems are characterized by heterocoordination (Table 6), which is stronger in the solution with 1,4-dimethylbenzene. Calculations in terms of the Kirkwood-Buff integrals formalism [81] show the existence of unlike aggregates at higher mole factions of $C_6F_6$ in the solution with benzene [82].

*5.6 Mixtures with other fluorinated benzenes*

Firstly, we consider mixtures of the type $C_6H_{6-u}F_u$ ($u < 6$) + $C_6H_6$. Their $H_m^E$ values strongly depend on the number and position of the F atoms attached to the aromatic ring and not on the dipole moment of the AFC involved [26]. In spite of this, some general features can be stated. (i) The $H_m^E$ values are positive and not large [26]. The largest value is attained for the system with 1,3,5-$C_6H_3F_3$ (496 J mol$^{-1}$ [26]). (ii) The $C_{pm}^E$ values are negative ($-1.1$ J mol$^{-1}$ K$^{-1}$ for the mentioned solution [26]). That is, dispersive interactions are dominant. The $C_6F_6$ mixture behaves in different way and shows $H_m^E < 0$ and $C_{pm}^E > 0$ (see Tables 4 and 5). The $H_m^E$ curve of the system with $C_6HF_5$ is S-shaped with positive $H_m^E$ values at lower concentrations of the AFC and $C_{pm}^E = 0.16$ J mol$^{-1}$ K$^{-1}$ [26]. Regarding the $C_6HF_5 + C_6H_{12}$ mixture, its $H_m^E$ value (1527 J mol$^{-1}$ [83] is very close to that of the mixture with $C_6F_6$ (1576 J mol$^{-1}$[27]); $C_{pm}^E = -1.1$ J mol$^{-1}$ K$^{-1}$ [83] and the application of the Flory model gives $dev(H_m^E) = 0.028$. It is clear that, although $C_6HF_5$ is a polar compound ($\mu = 1.6$ D [19]), the system under consideration is mainly characterized by dispersive interactions. On the other hand, $V_m^E/cm^3$ mol$^{-1}$ values of the solutions with benzene (0.860) or with cyclohexane (2.082) are large and positive [84], which suggests that the contribution to $H_m^E$ from the eos term will be here also very important. Using $\alpha_{pi}$, and $\kappa_{T_i}$ values reported in reference [85] for $C_6HF_5$, we have determined $U_{Vm}^E$/J mol$^{-1}$ for the such solutions and the results are: $-253$ (benzene) and 827 (cyclohexane). We remark the negative value for the system with $C_6H_6$ that reveals that interactions between unlike molecules are dominant in that mixture. The S-shaped concentration dependence of $H_m^E$ has been explained assuming that the physical interactions are described by a term which is skewed to the region rich in the component of lower molar volume as it occurs within the regular solution theory [83]. However, $U_{V_m}^E$ values are negative over the entire concentration range and the S-shaped $H_m^E$ curve can be related to large eos contribution to this excess function. In addition, $V_m^E$/cm$^3$ mol$^{-1}$ values of the mixtures $C_6HF_5 + C_6H_6$ (0.860), or + $C_7H_8$ (0.505), or + 1,4-DMBZ (0.287) decrease when the number of the $CH_3$ groups attached to the aromatic ring increases [84]. The same trend is encountered for the corresponding systems with $C_6F_6$ (see above). This set of features reveals that a certain similarity exists between the systems with $C_6F_6$ or with $C_6HF_5$.

The DISQUAC application to this type of systems shows that the 1,3,5-$C_6H_3F_3$ + $C_6H_6$ mixture, characterized by one contact, can be described in terms of the DIS approximation. This is not the case for the systems $C_6HF_5$+ $C_6H_6$ or + $C_6H_{12}$ characterized by one and three contacts respectively (see above). Nevertheless, our calculations reveal that the coefficients $C_{bf,1}^{QUAC}$ and $C_{sf,1}^{QUAC}$ (s = b,c; l =1,3) of the mixtures with $C_6F_6$ can be used for the systems with $C_6HF_5$, while the $C_{sf,2}^{DIS/QUAC}$ ( s = b,c) coefficients must be modified (see Table S2 of supplementary material ). A similar trend has been encountered when treating in terms of DISQUAC mixtures such as, e.g., 1-alkanol + $n$-alkane [86], or + benzene, or + toluene [59], or + $N,N$-dialkylamide [57], or + pyridine or + alkyl-pyridine [87], or + DMSO [88]. It must be underlined that the $C_{cf,1}^{QUAC}$ (l =1,2,3) coefficients for solutions with $C_6H_5F$ or 1,4-$C_6H_4F_2$ are the same and that differ from those of mixtures with $C_6F_6$ or $C_6HF_5$, which suggests that the former type of systems behave differently.

*5.7 The aromacity effect*

$H_m^E$ values of $CH_3(CH_2)_uF$ + $n$-alkane mixtures decrease when $u$ is increased. Thus, $H_m^E$(hexane)/J mol$^{-1}$ = 481($u$ = 4); 416 ($u$ =5); 294 ($u$ = 7) [48]. This variation can be explained in terms of the effective dipole moments, $\bar{\mu}$, of 1-fluoroalkanes, an useful magnitude to examine the impact of polarity of bulk properties. It is defined by [49,76,89]:

$$\bar{\mu} = \left[\frac{\mu^2 N_A}{4\pi\varepsilon_0 V_m k_B T}\right]^{1/2} \tag{8}$$

where the symbols have the usual meaning. The dipole moments of 1-fluoroalkanes with $u$ = 4, or 5 are, respectively, (1.85 and 1.82) D, [90] and the corresponding molar volumes (cm$^3$ mol$^{-1}$) are, in the same order, 114.90 [91] and 130.03 (at 293.15 K) [92]. The $\bar{\mu}$ values are: 0.660 ($u = 4$) and 0.611 ($u = 5$). The observed decrease of $H_m^E$ can be then ascribed to a slight weakening of the dipolar interactions between fluoroalkane molecules when $u$ is increased. The $H_m^E$ value for the 1-fluorohexane system is much lower than the corresponding result for the solution with fluorobenzene (869 J mol$^{-1}$ [17]) and $H_{m,1}^{E,\infty}$ (hexane)/kJ mol$^{-1}$ = 2.1 ($u$ = 5 [48]) < 3.4 (C$_6$H$_5$F) [17]. Since $\bar{\mu}$ (fluorobenzene) = 0.664 is quite similar to the values reported above for 1-fluoroalkanes, and taking into account the lack of volumetric data for these solutions needed to $U_{Vm}^E$ determinations, the large difference between these enthalpic data can be explained assuming that more dispersive interactions between fluorobenzene molecules are broken upon mixing (aromacity effect).

On the other hand, for benzene systems, $H_m^E$/J mol$^{-1}$ = 98 ($u$ = 4); 133 ($u$ =5); 219 ($u$ = 7) [48]. Therefore, $H_m^E$ increases in line with $u$, which can be ascribed to the longer 1-fluoroalkanes can break more easily benzene-benzene interactions due to their larger aliphatic surface. These values are lower than the corresponding results for hexane mixtures, and reveal the existence of interactions between unlike molecules. In the case of the C$_6$H$_5$F + C$_6$H$_6$ mixture, both components have rather similar aromatic surfaces and $H_m^E$ is 1 J mol$^{-1}$ (at 303.15 K) [51].

### 6. Conclusions

Mixtures containing fluorobenzene, or 1,4-difluorobenzene or hexafluorobenzene and alkane or an aromatic hydrocarbon have been investigated on the basis of thermodynamic properties from the literature, and using the DISQUAC and UNIFAC (Dortmund) models and the $S_{CC}(0)$ formalism. It has been shown that mixtures with alkanes: (i) are characterized by interactions between like molecules; (ii) that these interactions are essentially dispersive; (iii) that the Patterson's effect does not exist in mixtures with *n*-alkanes; (iv) structural effects can be very important. In fact, for mixtures with a given *n*-alkane, $H_m^E$ values change in the sequence C$_6$F$_6$ > 1,4-C$_6$H$_4$F$_2$ > C$_6$H$_6$ > C$_6$H$_5$F, while $U_{Vm}^E$ changes in the order: 1,4-C$_6$H$_4$F$_2$ > C$_6$H$_6$ ≈ C$_6$H$_5$F > C$_6$F$_6$. On the other hand, C$_6$F$_6$ + aromatic hydrocarbon mixtures are characterized by interactions between unlike molecules. The application of $S_{CC}(0)$ formalism reveals that homocoordination is more important in C$_6$F$_6$ + *n*-alkane mixtures than in the corresponding systems with C$_6$H$_5$F, and that heterocoordination is dominant in the solutions of C$_6$F$_6$ with an aromatic hydrocarbon. In systems with *n*-alkanes, the aromacity effect leads to increased $H_m^E$ values for mixtures with C$_6$H$_5$F in comparison to those of systems with 1-fluoroalkanes.

**Funding**

This work was supported by Consejería de Educación de Castilla y León, under Project VA100G19 (Apoyo a GIR, BDNS: 425389).

**TABLE 1**

Dispersive (DIS) and quasichemical (QUAC) interchange coefficients, $C_{\text{sf},l}^{\text{DIS}}$ and $C_{\text{sf},l}^{\text{QUAC}}$, for (s,f,) contacts[a] in fluorobenzene, or 1,4-difluorobenzene, or hexafluorobenzene + hydrocarbon mixtures ($l = 1$, Gibbs energy; $l = 2$, enthalpy; $l = 3$, heat capacity).

| System | (s,f) | $C_{\text{sf},1}^{\text{DIS}}$ | $C_{\text{sf},2}^{\text{DIS}}$ | $C_{\text{sf},3}^{\text{DIS}}$ | $C_{\text{sf},1}^{\text{QUAC}}$ | $C_{\text{sf},2}^{\text{QUAC}}$ | $C_{\text{sf},3}^{\text{QUAC}}$ |
|---|---|---|---|---|---|---|---|
| $C_6H_5F$ + $n$-alkane[b] | (a,f) | −1.7 | −3 | −0.82[c] | 2 | 3.8 | 0.2 |
| 1,4-$C_6H_4F_2$ + $n$-alkane[b] | (a,f) | −1.7 | −2.3 | −0.82 | 2 | 3.8 | 0.2 |
| $C_6F_6$ + $n$-alkane | (a,f) | 1.035 | 0.32 | 2.9[d] | −0.57 | 0.35 | −3.06 |
| $C_6H_5F$ + $C_6H_{12}$[e] | (c,f) | −1.55 | −2.65 | −0.1 | 2 | 3.8 | 0.2 |
| 1,4-$C_6H_4F_2$ + $C_6H_{12}$[e] | (c,f) | −1.55 | −2.65 | −0.1 | 2 | 3.8 | 0.2 |
| $C_6F_6$ + $C_6H_{12}$ | (c,f) | 1.25 | 0.83 | 2.7 | −0.57 | 0.35 | −3.06 |
| $C_6F_6$ + $C_6H_6$ | (b,f) | 2.02 | 2.25 | −4.4 | −1.8 | −2.2 | 4 |
| $C_6F_6$ + $C_7H_8$ | (b,f) | 1.46 | 0.4 | 1 | −1.8 | −2.2 | 4 |
| $C_6F_6$ + 1,4-$C_6H_4(CH_3)_2$ | (b,f) | 0.3 | −2.32 | 1 | −1.8 | −2.2 | 4 |

[a] type s = a, aliphatic, s = b, aromatic, s = c, cyclic; type f, fluorine; [b] $C_{\text{af},l}^{\text{DIS/QUAC}}$ coefficients determined assuming that $C_{\text{bf},l}^{\text{DIS/QUAC}} = 0$ ($l =1,2,3$); [c] $C_{\text{af},3}^{\text{DIS}} = -1.0$ ($n$-$C_6$); −0.82 ($n$-$C_7$); −0.58 ($n$-$C_8$); −0.5 ($n$-$C_{10}$); −0.2 ($n$-$C_{12}$); 0.1($\geq n$-$C_{14}$); [d] value for heptane; for ($n$-$C_{14}$), $C_{\text{af},3}^{\text{DIS}} = 2.78$; [e] $C_{\text{cf},l}^{\text{DIS/QUAC}}$ coefficients determined assuming that $C_{\text{bf},l}^{\text{DIS/QUAC}} = 0$ ($l = 1,2,3$) (see text)

**TABLE 2**

Molar excess Gibbs energy, $G_\text{m}^\text{E}$, at equimolar composition and temperature $T$, for fluorobenzene(1) or hexafluorobenzene(1) + hydrocarbon(2) mixtures. Comparison of experimental results (Exp.) with DISQUAC (DQ) calculations using the interaction parameters from Table 1, or with UNIFAC (Dortmund) (UNIF) predictions obtained using interaction parameters from the literature [9,18].

| Hydrocarbon | $T$/K | $N^\text{a}$ | $G_\text{m}^\text{E}$/J mol$^{-1}$ | | $\sigma_\text{r}(P)^\text{b}$ | | | Ref |
|---|---|---|---|---|---|---|---|---|
| | | | Exp$^\text{c}$ | DQ. | Exp$^\text{c}$ | DQ$^\text{d}$ | UNIF$^\text{e}$ | |
| C$_6$H$_5$F + alkane | | | | | | | | |
| $n$-C$_6$ ($x_1$ = 0.537) | 344.9 | | 393 | 380 | | | | 93 |
| | | | | 377$^\text{f}$ | | | | |
| $n$-C$_7$ ($x_1$ = 0.491) | 359.9 | | 330 | 346 | | | | 93 |
| | | | | 290$^\text{f}$ | | | | |
| CH$_3$-C$_6$H$_{12}$ | 348.15 | 25 | 357 | 353 | 0.0005 | 0.001 | 0.015 | 94 |
| C$_6$H$_{12}$ | 323.15 | 10 | 341 | 419 | 0.011 | 0.028 | 0.014 | 72 |
| | 343.15 | 10 | 357 | 395 | 0.016 | 0.025 | 0.026 | 72 |
| | 348.15 | 28 | 387 | 389 | 0.0005 | 0.004 | 0.025 | 95 |
| C$_6$F$_6$ + hydrocarbon | | | | | | | | |
| $n$-C$_{14}$ | 298.15 | 20 | 641 | 646 | 0.005 | 0.018 | 0.022 | 62 |
| $n$-C$_{16}$ | 298.15 | 18 | 604 | 615 | 0.003 | 0.023 | 0.031 | 63 |
| C$_6$H$_{12}$ | 303.15 | 10 | 797 | 800 | 0.0035 | 0.004 | 0.067 | 96 |
| | 333.15 | 10 | 720 | 707 | 0.0007 | 0.003 | 0.059 | 96 |
| | 343.15 | 10 | 698 | 703 | 0.0013 | 0.003 | 0.057 | 96 |
| C$_6$H$_6$ | 303.15 | 10 | −59 | −53 | 0.0003 | 0.009 | 0.005 | 24 |
| | 313.15 | 10 | −44 | −39 | 0.0002 | 0.008 | 0.002 | 24 |
| | 333.15 | 10 | −18 | −16 | 0.0014 | 0.007 | 0.004 | 24 |
| | 343.15 | 10 | −4.5 | −5.3 | 0.002 | 0.006 | 0.004 | 24 |
| Toluene | 303.15 | 10 | −213 | −212 | 0.0005 | 0.005 | 0.005 | 24 |
| | 343.15 | 10 | −106 | −110 | 0.0006 | 0.003 | 0.016 | 24 |
| 14DMBZ$^\text{g}$ | 303.15 | 9 | −430 | −445 | 0.0006 | 0.009 | 0.021 | 24 |
| | 323.15 | 9 | −353 | −365 | 0.0021 | 0.005 | 0.046 | 24 |
| | 343.15 | 9 | −305 | −296 | 0.0065 | 0.005 | 0.008 | 24 |

$^\text{a}$number of data points; $^\text{b}$equation (5); $^\text{c}$values of the fittings of experimental data using Redlich-Kister expansions to represent $G_\text{m}^\text{E}$; $^\text{d}$results from equation (5) obtained using DISQUAC values for $P_\text{calc}$; $^\text{e}$results from equation (5) obtained using UNIFAC values for $P_\text{calc}$; $^\text{f}$UNIFAC result; $^\text{g}$1,4-dimethylbenzene

**TABLE 3**

Azeotropic data ($x_{1az}, T_{az}; P_{az}$) for fluorobenzene(1) or hexafluorobenzene(1) + hydrocarbon(2) mixtures. Comparison of experimental results (Exp.) with DISQUAC (DQ) calculations using the interaction parameters from Table 1

| System | $T_{az}$/K | $x_{1az}$ | | $P_{az}$/kPa | | Ref |
|---|---|---|---|---|---|---|
| | | Exp | DQ. | Exp. | DQ. | |
| $C_6H_5F + n\text{-}C_6$ | 342 | 0.046[a] | 0.002 | 101.4 | 101.1 | 93 |
| $C_6H_5F + n\text{-}C_7$ | 357.6 | 0.896[a] | 0.898 | 101.4 | 101.2 | 93 |
| $C_6F_6 + n\text{-}C_6$ | 341.09 | 0.245[a] | 0.261 | 101.3 | 105.0 | 97 |
| $C_6F_6 + CH_3\text{-}C_6H_{11}$ | 353.02 | 0.949[a] | 0.976 | 101.3 | 100.8 | 97 |
| | 343.15 | 0.468[a] | 0.463 | 93.21 | 93.43 | 96 |
| $C_6F_6 + C_6H_6$ | 302.6 | 0.047[a] | | 16.00 | | 23 |
| | 303.15 | 0.048[a] | 0.088 | 15.93 | 16.1 | 24 |
| | 305.05 | 0.8115[b] | n.a.[c] | 16.00 | | 23 |
| | 333.15 | 0.126[a] | 0.143 | 52.64 | 53.2 | 24 |
| | 343.15 | 0.157[a] | 0.155 | 74.40 | 75.1 | 24 |
| | 352.45 | 0.208[a] | | 101.3 | | 97 |
| | 353.45 | | 0.182[a] | | 104.7 | |
| | 353.45 | 0.813[b] | 0.700 | 101.3 | 101.7 | 97 |
| | 357.8 | 0.193[a] | | 120. | | 23 |
| | 358.85 | | 0.188[a] | | 120.6 | |
| | 358.85 | 0.774[b] | 0.619 | 120 | 118.4 | 23 |

[a]positive azeotrope; [b]negative azeotrope; [c]not available: DISQUAC does not predict the existence of negative azeotropes for this system at low temperatures (see Text)

**TABLE 4**

Molar excess enthalpies, $H_m^E$, at equimolar composition and temperature $T$, for fluorobenzene(1), or 1,4-difluorobenzene(1) or hexafluorobenzene(1) + hydrocarbon(2) mixtures. Comparison of experimental results (Exp.) with DISQUAC (DQ) calculations using the interaction parameters from Table 1.

| hydrocarbon | T/K | $N^a$ | $H_m^E$/J mol$^{-1}$ | | $dev(H_m^E)^b$ | | | Ref |
|---|---|---|---|---|---|---|---|---|
| | | | Exp$^c$ | DQ. | Exp$^c$ | DQ$^d$ | UNIF$^e$ | |
| $C_6H_5F$ + alkane | | | | | | | | |
| n-C$_6$ | 298.15 | 15 | 869 | 845 | 0.003 | 0.020 | 0.028 | 17 |
| n-C$_7$ | 298.15 | 15 | 888 | 901 | 0.001 | 0.012 | 0.005 | 17 |
| n-C$_8$ | 298.15 | 17 | 941 | 950 | 0.003 | 0.010 | 0.005 | 17 |
| n-C$_{10}$ | 298.15 | 18 | 1024 | 1034 | 0.0008 | 0.011 | 0.006 | 17 |
| n-C$_{12}$ | 298.15 | 18 | 1092 | 1102 | 0.001 | 0.010 | 0.006 | 17 |
| n-C$_{14}$ | 298.15 | 18 | 1156 | 1159 | 0.003 | 0.005 | 0.008 | 17 |
| n-C$_{16}$ | 298.15 | 18 | 1201 | 1208 | 0.005 | 0.006 | 0.006 | 17 |
| C$_6$H$_{12}$ | 298.15 | 21 | 864 | 878 | 0.001 | 0.016 | 0.059 | 71 |
| | 303.15 | 8 | 906 | 867 | 0.007 | 0.041 | 0.031 | 72 |
| | 313.15 | 9 | 885 | 849 | 0.024 | 0.037 | 0.064 | 72 |
| CH$_3$-C$_6$H$_{11}$ | 283.15 | | 790 | 823 | | | | 70 |
| | 298.15 | | 738 | 809 | | | | 70 |
| | | 21 | 823 | | 0.001 | 0.018 | 0.063 | 71 |
| | 313.15 | | 692 | 793 | | | | 70 |
| 1,4-difluorobenzene + n-alkane | | | | | | | | |
| n-C$_7$ | 298.15 | 12 | 1050 | 1019 | 0.007 | 0.020 | 0.025 | 15 |
| n-C$_8$ | 298.15 | 13 | 1083 | 1071 | 0.009 | 0.025 | 0.025 | 15 |
| n-C$_{10}$ | 298.15 | 11 | 1185 | 1158 | 0.005 | 0.025 | 0.015 | 15 |
| n-C$_{12}$ | 298.15 | 14 | 1290 | 1228 | 0.010 | 0.034 | 0.021 | 15 |
| n-C$_{16}$ | 298.15 | 13 | 1359 | 1334 | 0.017 | 0.025 | 0.015 | 15 |
| C$_6$F$_6$ + hydrocarbon | | | | | | | | |
| n-C$_7$ | 298.15 | 28 | 1119 | 1098 | 0.003 | 0.020 | 0.253 | 15 |
| n-C$_{14}$ | 298.15 | 24 | 1324 | 1400 | 0.003 | 0.049 | 0.361 | 15 |
| C$_6$H$_{12}$ | 298.15 | 9 | 1576 | 1577 | 0.006 | 0.016 | 0.167 | 27 |
| | 313.2 | 9 | 1517 | 1549 | 0.015 | 0.014 | 0.111 | 27 |
| | 333.2 | 5 | 1483 | 1509 | 0.004 | 0.011 | 0.094 | 27 |
| C$_6$H$_{11}$-CH$_3$ | 298.15 | | 1315 | 1275 | | | | 27 |

Table 4 (continued)

| | T/K | n[a] | $H_{m,exp}^E$[b] | $H_{m,calc}^E$[c] | $\sigma_r$(H^E_m)[c] | $\sigma_r$(H^E_m)[d] | $\sigma_r$(H^E_m)[e] | Ref. |
|---|---|---|---|---|---|---|---|---|
| | 333.2 | | 1239 | 1235 | | | | 27 |
| C$_6$H$_6$ | 283.2 | | −447 | −581 | | | | 26 |
| | 298.2 | 12 | −456 | −507 | 0.007 | 0.081 | 0.621 | 27 |
| | | | −496 | | | | | 26 |
| | 308.15 | | −464 | −465 | | | | 26 |
| | 313.2 | 13 | −433 | −446 | 0.009 | | | 27 |
| | 318.15 | | −439 | −429 | | | | 26 |
| | 328.2 | 10 | −402 | −396 | 0.005 | | | 27 |
| | 343.9 | 9 | −351 | −352 | 0.006 | 0.117 | 0.641 | 27 |
| Toluene | 313.2 | | −1140 | −1118 | | | | 27 |
| | 328.2 | | −908 | −906 | | | | 27 |
| | 343.2 | | −651 | −702 | | | | 27 |
| 14DMBZ[f] | 313.2 | 10 | −1660 | −1660 | 0.006 | 0.037 | 0.270 | 27 |
| | 328.2 | 7 | −1504 | −1516 | 0.009 | 0.040 | 0.176 | 27 |
| | 343.2 | 8 | −1430 | −1379 | 0.021 | 0.041 | 0.087 | 27 |

[a]number of data points; [b]equation (6); [c]results from equation (6) obtained using $H_{m,calc}^E$ values from the fitting of experimental data to Redlich-Kister expansions; [d]results from equation (6) obtained using $H_{m,calc}^E$ values from the DISQUAC application; [e]results from equation (6) obtained using $H_{m,calc}^E$ values from the UNIFAC application using interaction parameters from the literature [9,18].
[f]1,4-dimethylbenzene

**TABLE 5**

Isobaric molar excess heat capacities, $C_{pm}^{E}$, of fluorobenzene(1), or hexafluorobenzene(1) + hydrocarbon(2) mixtures at 298.15 K and equimolar composition. Comparison of experimental results (Exp.) with DISQUAC (DQ) calculations using the interaction parameters from Table 1, or with UNIFAC (Dortmund) predictions obtained using interaction parameters from the literature [9,18].

| System | $C_{pm}^{E}$/J mol$^{-1}$ K$^{-1}$ | | | Ref. |
|---|---|---|---|---|
| | Exp | DQ | UNIF | |
| $C_6H_5F$ + $n$-$C_7$ | −1.18 | −1.20 | 0.16 | 17 |
| $C_6H_5F$ + $n$-$C_8$ | −1.42 | −1.45 | 0.19 | 17 |
| $C_6H_5F$ + $n$-$C_{10}$ | −2.00 | −2.00 | 0.26 | 17 |
| $C_6H_5F$ + $n$-$C_{12}$ | −2.78 | −2.74 | 0.33 | 17 |
| $C_6H_5F$ + $n$-$C_{14}$ | −3.75 | −3.66 | 0.39 | 17 |
| $C_6F_6$ + $n$-$C_7$ | −1.18 | −1.03 | 12.6 | 15 |
| $C_6F_6$ + $n$-$C_{14}$ | −2.25 | −2.13 | 13.9 | 15 |
| $C_6F_6$ + $C_6H_{12}$ | −2.7 | −1.75 | 0.77 | 27 |
| $C_6F_6$ + $C_6H_6$ | 1.8 | 4.4 | −1.6 | 27 |
| $C_6F_6$ + $C_7H_8$ | 16.7 | 15.1 | 0.56 | 27 |
| $C_6F_6$ +14DMBZ[a] | 12.3 | 10.5 | 1.8 | 27 |

[a]14DMBZ, 1,4-dimethylbenzene

**TABLE 6**

$S_{cc}(0)$ of fluorobenzene(1), or hexafluorobenzene(1), or benzene(1) + hydrocarbon(2) mixtures at temperature $T$ and at composition $x_1$, where $S_{cc}(0)$ shows the maximum value. Comparison of experimental results (Exp.) with DISQUAC values obtained using interaction parameters from Table 1.

| System | $T$/K | Exp. | | DQ | | Ref. |
|---|---|---|---|---|---|---|
| | | $x_1$ | $S_{cc}(0)$ | $x_1$ | $S_{cc}(0)$ | |
| $C_6H_5F$ + $n$-$C_7$ | 298.15 | | | 0.54 | 0.378 | |
| $C_6H_5F$ + $n$-$C_{14}$ | 298.15 | | | 0.58 | 0.305 | |
| $C_6H_5F$ + $n$-$C_{16}$ | 298.15 | | | 0.57 | 0.284 | |
| $C_6H_5F$ + $C_6H_{12}$ | 298.15 | | | 0.50 | 0.389 | |
| $C_6F_6$ + $n$-$C_7$ | 298.15 | | | 0.51 | 0.528 | |
| $C_6F_6$ + $n$-$C_{14}$ | 298.15 | 0.61 | 0.458 | 0.66 | 0.559 | 62 |
| $C_6F_6$ + $n$-$C_{16}$ | 298.15 | 0.63 | 0.458 | 0.69 | 0.542 | 63 |
| $C_6F_6$ + TMP | 298.15 | 0.50 | 0.374 | | | 79 |
| $C_6F_6$ + $C_6H_{12}$ | 303.15 | 0.44 | 0.704 | 0.44 | 0.711 | 96 |
| $C_6F_6$ + $C_6H_6$ | 303.15 | 0.50 | 0.231 | 0.50 | 0.219 | 24 |
| $C_6F_6$ + $C_7H_8$ | 303.15 | 0.50 | 0.206 | 0.50 | 0.200 | 24 |
| $C_6F_6$ + 14DMBZ[a] | 303.15 | 0.50 | 0.177 | 0.50 | 0.177 | 24 |
| $C_6H_6$ + $n$-$C_7$ | 298.15 | 0.50 | 0.354 | | | 68 |
| $C_6H_6$ + $n$-$C_{14}$ | 298.15 | 0.65 | 0.291 | | | 98 |
| $C_6H_6$ + $n$-$C_{16}$ | 298.15 | 0.57 | 0.256 | | | 99 |
| $C_6H_6$ + TMP[b] | 308.15 | 0.65 | 0.308 | | | 100 |

[a] 14DMBZ, 1,4-dimethylbenzene; [b]TMP, 2,24-trimethylpentane

**TTABLE 7**

Excess molar functions: enthalpy, $H_m^E$, volume, $V_m^E$, and internal energy at constant volume $U_{Vm}^E$, at equimolar composition and 298.15 K for $C_6H_5F$, or 1,4-$C_6H_4F_2$, or $C_6F_6$, or benzene + organic solvent (Solv) mixtures at 298.15 K

| Solv | $C_6H_5F$ | | | 1,4-$C_6H_4F_2$ | | | $C_6F_6$ | | | $C_6H_6$ | | |
|---|---|---|---|---|---|---|---|---|---|---|---|---|
| | $H_m^{E\ a}$ | $V_m^{E\ b}$ | $U_{Vm}^{E\ a}$ | $H_m^{E\ a}$ | $V_m^{E\ b}$ | $U_{Vm}^{E\ a}$ | $H_m^{E\ a}$ | $V_m^{E\ b}$ | $U_{Vm}^{E\ a}$ | $H_m^{E\ a}$ | $V_m^{E\ b}$ | $U_{Vm}^{E\ a}$ |
| $n$-$C_6$ | 869[c] | 0.246[c] | 803 | | | | | | | 924[i] | 0.403[m] | 816 |
| $n$-$C_7$ | 888[c] | 0.431[c] | 765 | 1050[e] | 0.650[e] | 854 | 1119[e] | 1.882[e] | 557 | 931[j] | 0.562[m] | 768 |
| $n$-$C_8$ | 941[c] | 0.532[c] | 785 | 1083[e] | 0.779[e] | 842 | | | | 969[k] | 0.700[m] | 761 |
| $n$-$C_{10}$ | 1024[c] | 0.664[c] | 822 | 1185[e] | 0.935[e] | 886 | | | | | | |
| $n$-$C_{12}$ | 1092[c] | 0.740[c] | 863 | 1290[e] | 0.985[e] | 957 | | | | 1101[k] | 0.919[n] | 814 |
| $n$-$C_{14}$ | 1156[c] | 0.789[c] | 905 | | | | 1330[e] | 2.093[e] | 640 | 1183[k] | 1.015[o] | 856 |
| $n$-$C_{16}$ | 1201[c] | 0.832[c] | 935 | 1359[e] | 1.080[e] | 999 | | | | 1256[k] | 1.023[n] | 926 |
| $C_6H_{12}$ | 864[d] | 0.686[d] | 632 | | | | 1576[f] | 2.454[g] | 729 | 800[l] | 0.654[p] | 576 |
| $C_6H_6$ | | | | | | | −433[f,*] | 0.801[h,*] | −719[*] | | | |
| $C_7H_8$ | | | | | | | −1140[f,*] | 0.416[h,*] | −1298[*] | | | |
| 14DMBZ[q] | | | | | | | −1660[f,*] | 0.0855[h,*] | −1690[*] | | | |

[a]units: J mol$^{-1}$; [b]units: cm$^3$ mol$^{-1}$; [c][17]; [d][71]; [e][15]; [f][26]; [g][73], [h][75]; [i][101]; [j][10]; [k][11]; [l][102]; [m][78]; [n][103]; [o][104]; [p][71]; [q]1,4-dimethylbenzene; *, values at 313.15 K.

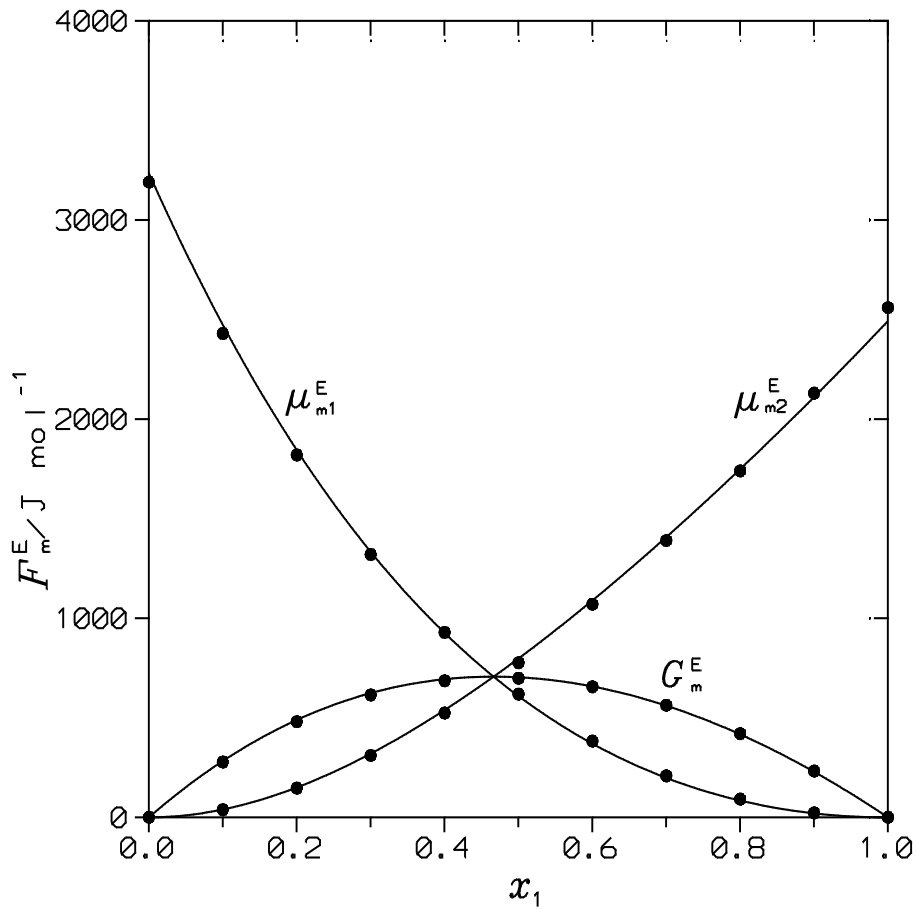

**Figure 1** Excess molar Gibbs energy, $F_m^E = G_m^E$, and excess molar chemical potentials, $F_m^E = \mu_{mi}^E$ of hexafluorobenzene(1) + cyclohexane(2) mixture at 343.15 K. Points, experimental results [96]. Solid lines, DISQUAC calculations.

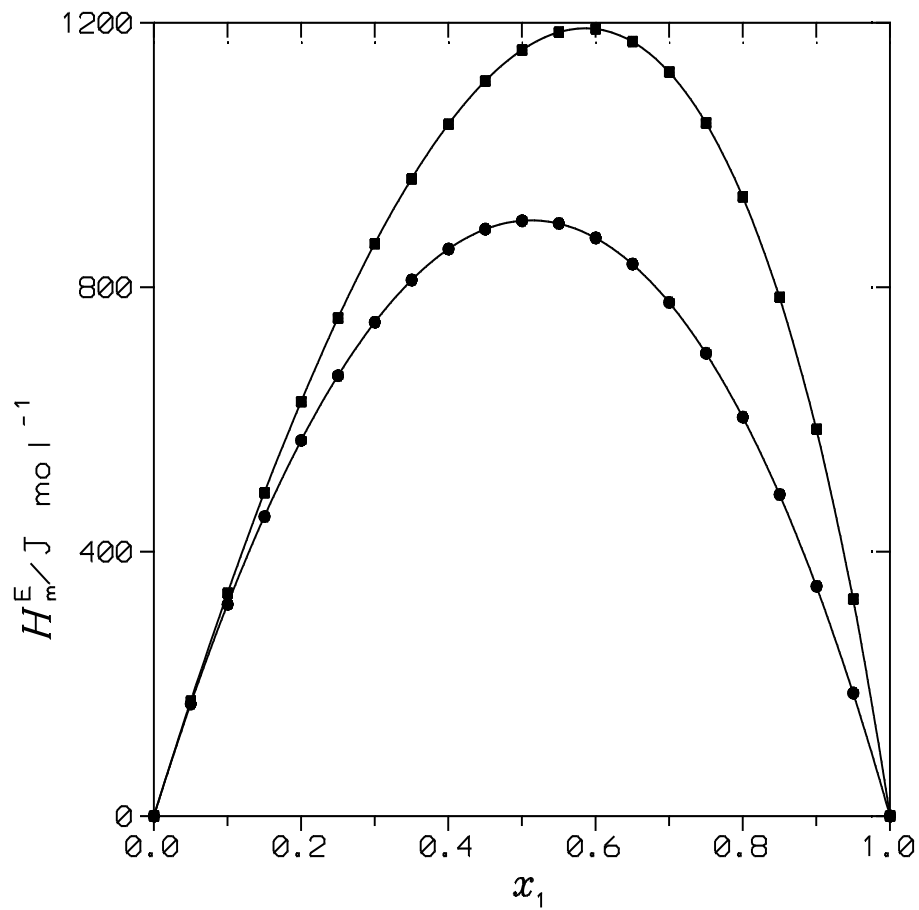

**Figure 2.** $H_m^E$ of fluorobenzene(1) + $n$-alkane(2) mixtures at 298.15 K. Points, experimental results [17]: (●), heptane; (■), tetradecane. Solid lines, DISQUAC calculation

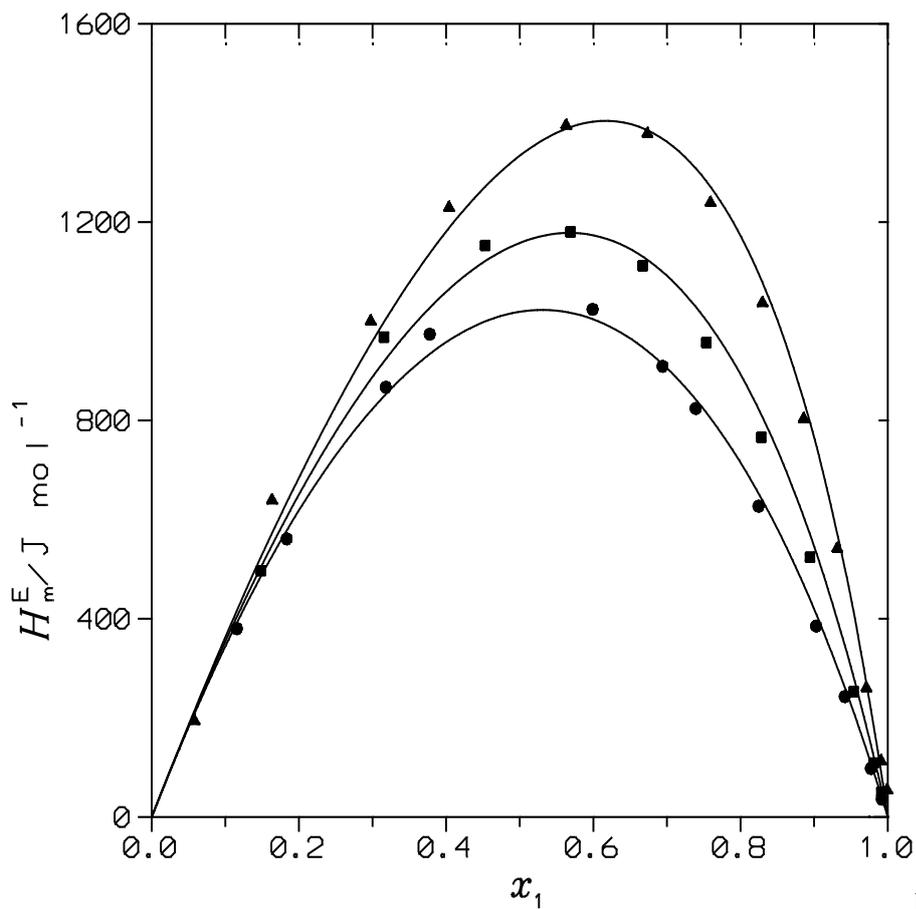

**Figure 3.** $H_m^E$ of 1,4-difluorobenzene(1) + *n*-alkane(2) mixtures at 298.15 K. Points, experimental results [15]: (●), heptane; (■), decane; (▲), tetradecane. Solid lines, DISQUAC calculations.

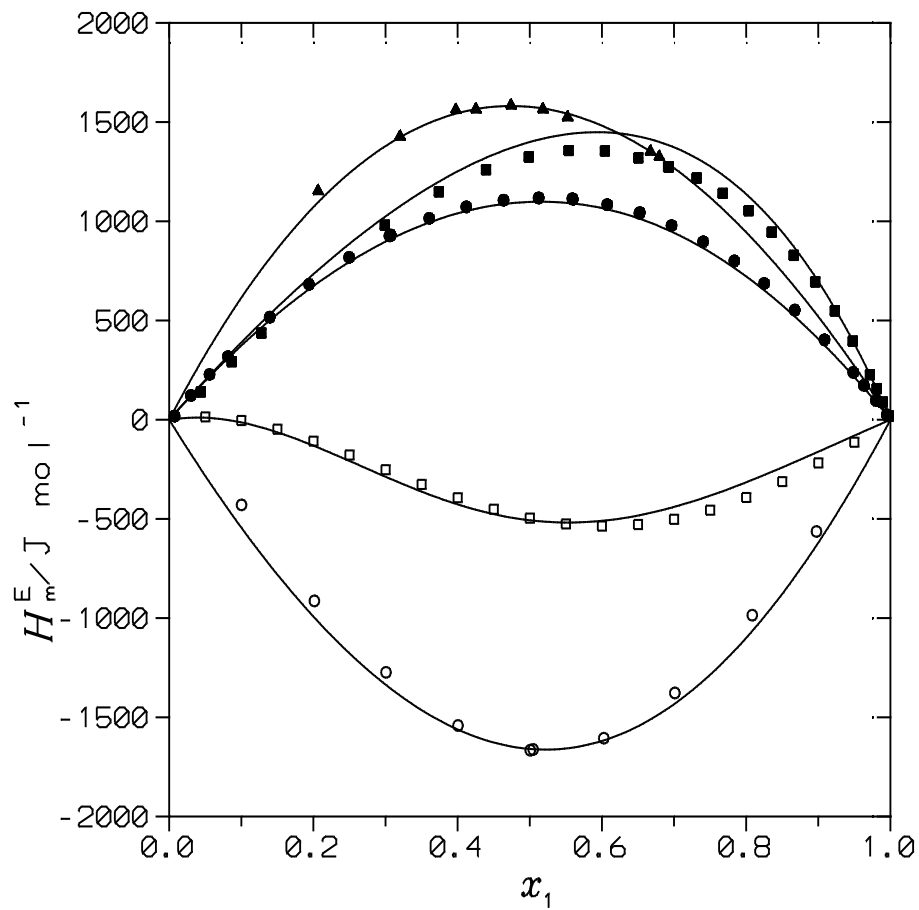

**Figure 4.** $H_m^E$ of hexafluorobenzene(1) + hydrocarbon(2) mixtures. Points, experimental results: (●), heptane [15]; (■), tetradecane [15]; (▲), cyclohexane [27]; (□), benzene [27] ($T$ = 298.15 K); (O), 1,4-dimethylbenzene [27] ($T$ = 313.15 K). Solid lines, DISQUAC calculations.

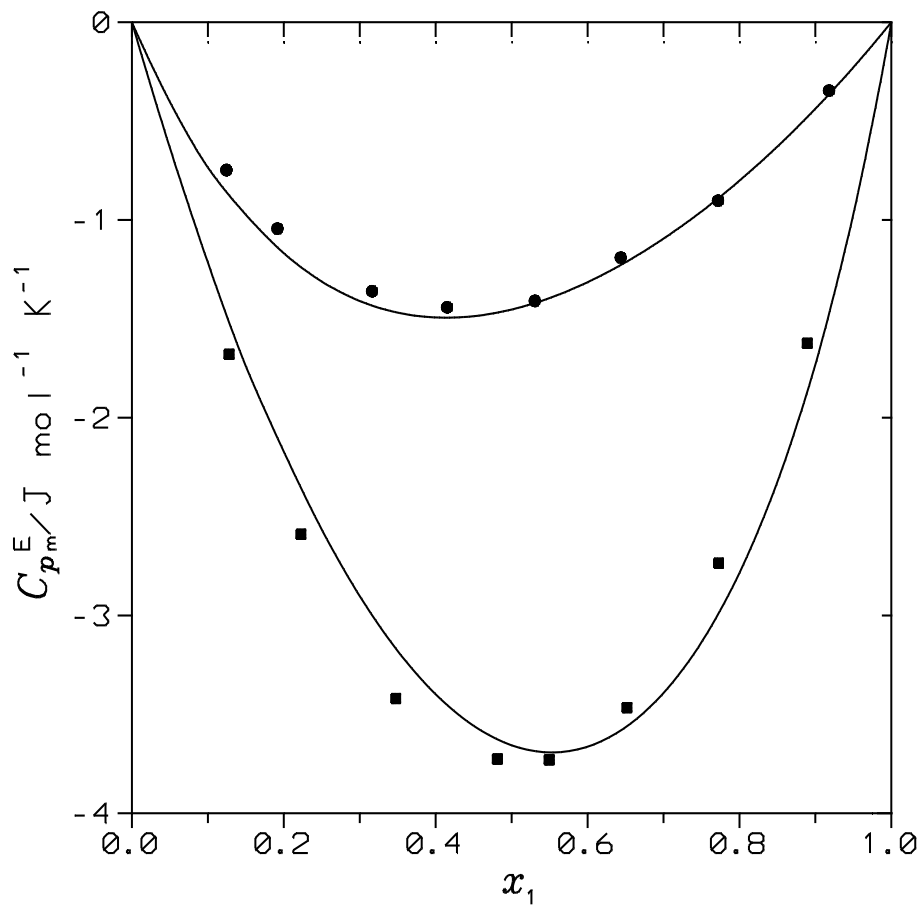

**Figure 5.** $C_{pm}^{E}$ of fluorobenzene(1) + $n$-alkane(2) mixtures at 298.15 K. Points, experimental results [17]: (●), heptane; (■), tetradecane. Solid lines, DISQUAC calculations.

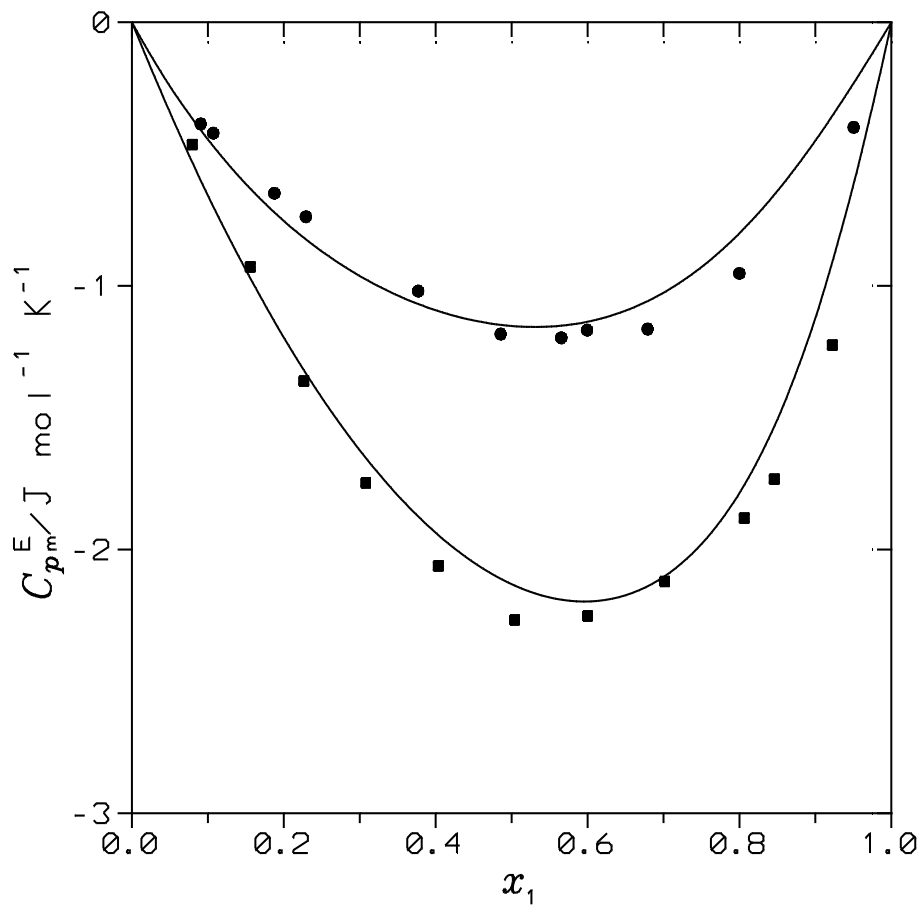

**Figure 6.** $C_{pm}^E$ of hexafluorobenzene(1) + *n*-alkane(2) mixtures at 298.15 K. Points, experimental results [15]: (●), octane; (■), tetradecane. Solid lines, DISQUAC calculations.

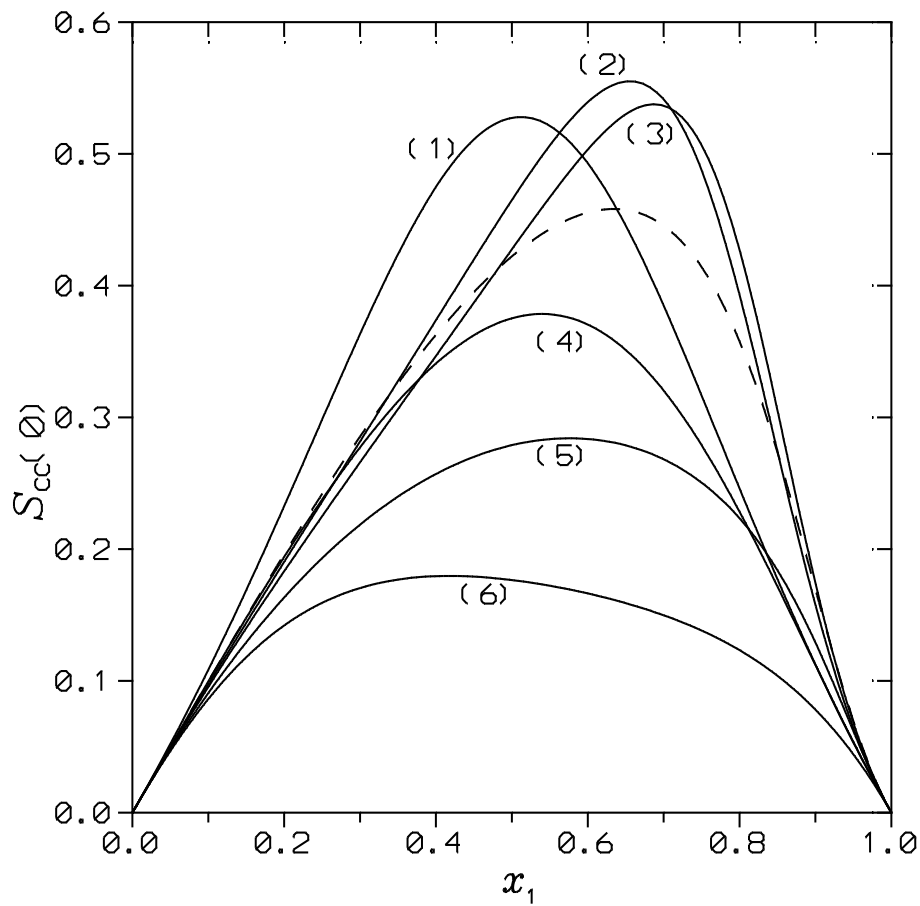

**Figure 7.** $S_{CC}(0)$ of fluorobenzene(1), or hexafluorobenzene(1) + hydrocarbon(2) mixtures at 298.15 K. Solid lines, DISQUAC calculations: (1) hexafluorobenzene(1) + heptane(2); (2), hexafluorobenzene(1) + tetradecane(2); (3), hexafluorobenzene(1) + hexadecane(2); (4), fluorobenzene(1)+ hexadecane(2); (5) fluorobenzene(1) + heptane(2); (6), hexafluorobenzene(1) + benzene(2). Dashed line, experimental results for hexafluorobenzene(1) + hexadecane(2) [63].

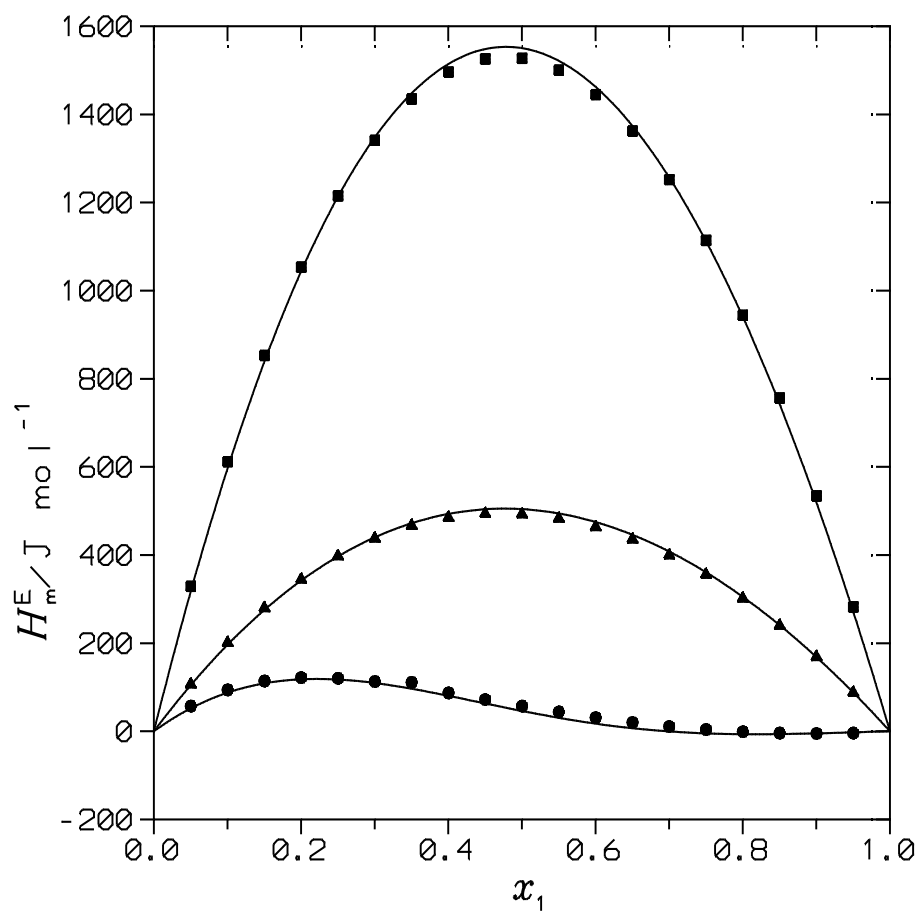

**Figure 8.** $H_m^E$ of fluorinated benzene(1) + hydrocarbon(2) mixtures at 298.15 K. Points, experimental results: (●), $C_6HF_5$ + benzene [26]; (▲), 1,3,5-trifluorobenzene + benzene [26]; (■), $C_6HF_5$ + cyclohexane [83]. Solid lines, DISQUAC calculations with interaction parameters from Table S2.

**SUPPLEMENTARY MATERIAL**

**THERMODYNAMICS OF MIXTURES CONTAINING A FLUORINATED BENZENE AND A HYDROCARBON**

Juan Antonio González,[a*] Luis Felipe Sanz,[a] Fernando Hevia,[b] Isaías García de la Fuente,[a] and José Carlos Cobos[a]

[a]G.E.T.E.F., Departamento de Física Aplicada, Facultad de Ciencias, Universidad de Valladolid, Paseo de Belén, 7, 47011 Valladolid, Spain.

[b]Université Clermont Auvergne, CNRS. Institut de Chimie de Clermont-Ferrand. F-63000, Clermont-Ferrand, France b Departamento de Física Aplacada. EIFAB. Campus D

*corresponding author, e-mail: jagl@termo.uva.es; Fax: +34-983-423136; Tel: +34-983-423757

**TABLE S1**

Physical properties, molar volume, $V_m$, isobaric expansion coefficient, $\alpha_p$ and isothermal compressibility $\kappa_T$ of aromatic compounds considered in this work.

| Compound | $V_m$ / cm³·mol⁻¹ | $\alpha_p$ /($10^{-3}$ K⁻¹) | $\kappa_T$ /($10^{-12}$ Pa⁻¹) |
|---|---|---|---|
| $C_6H_5F$ | 96.10[a] | 1.18[b] | 957[b,c] |
| 1,4-$C_6H_4F_2$ | 98.11[d] | 1.24[d,e] | 890[f] |
| $C_6F_6$ | 115.80[g] | 1.41[g] | 1147[g] |
| $C_6H_6$ | 89.44[c] | 1.21[c] | 966[c] |
| $C_7H_8$ | 106.89[c] | 1.067[c] | 912[c] |
| 1,4-dimethylbenzene | 123.94[c] | 0.956[c] | 859[c] |

[a][17]; [b] [s1]; [c][s2]; [d][15]; [e][s3]; [f]estimated value; [g] [s4]

**TABLE S2**

Dispersive (DIS) and quasichemical (QUAC) interchange coefficients, $C_{sf,l}^{DIS}$ and $C_{sf,l}^{QUAC}$, for (s,f,) contacts[a] in 1,3-5-trifluorobenzene, or pentafluorobenzene + hydrocarbon mixtures ($l = 1$, Gibbs energy; $l = 2$, enthalpy; $l = 3$, heat capacity).

| System | (s,f) | $C_{sf,1}^{DIS}$ | $C_{sf,2}^{DIS}$ | $C_{sf,3}^{DIS}$ | $C_{sf,1}^{QUAC}$ | $C_{sf,2}^{QUAC}$ | $C_{sf,3}^{QUAC}$ |
|---|---|---|---|---|---|---|---|
| 1,3,5-$C_6H_3F_3$ + $C_6H_6$ | (b,f) | 0.7[b] | 1.74 | −2.2 | | | |
| $C_6HF_5$ + $C_6H_6$ | (b,f) | 2.02[b] | 3 | −4.4 | −1.8 | −1 | 4 |
| $C_6HF_5$ + $C_6H_{12}$ | (c,f) | 1.25[b] | 2.25 | 2.7 | −0.57 | −0.78 | −3.06 |

[a]type s = b, aromatic, s = c, cyclic; type f, fluorine; [b]guessed value

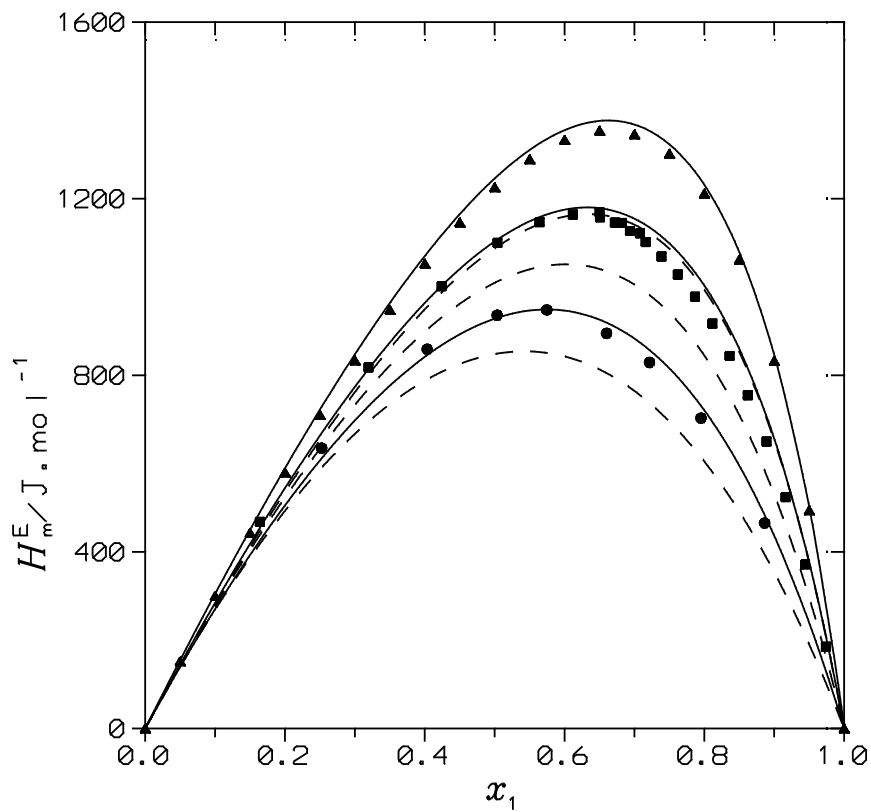

**Figure S1.** $H_m^E$ of benzene(1) + $n$-alkane(2) mixtures at 298.15 K. Points, experimental results: (●), heptane [10]; (■), dodecane [11]; (▲), hexadecane [11]. Solid lines, DISQUAC calculations using $C_{ab,2}^{DIS}$ = 0.559 (heptane); $C_{ab,2}^{DIS}$ = 0.575 (dodecane); $C_{ab,2}^{DIS}$ = 0.601 (hexadecane) [14]. Dashed lines, UNIFAC calculations with interaction parameters from the literature [9,18]: lower curve, results for the heptane solution; upper curve, results for the hexadecane mixture, between them, results for the dodecane system.

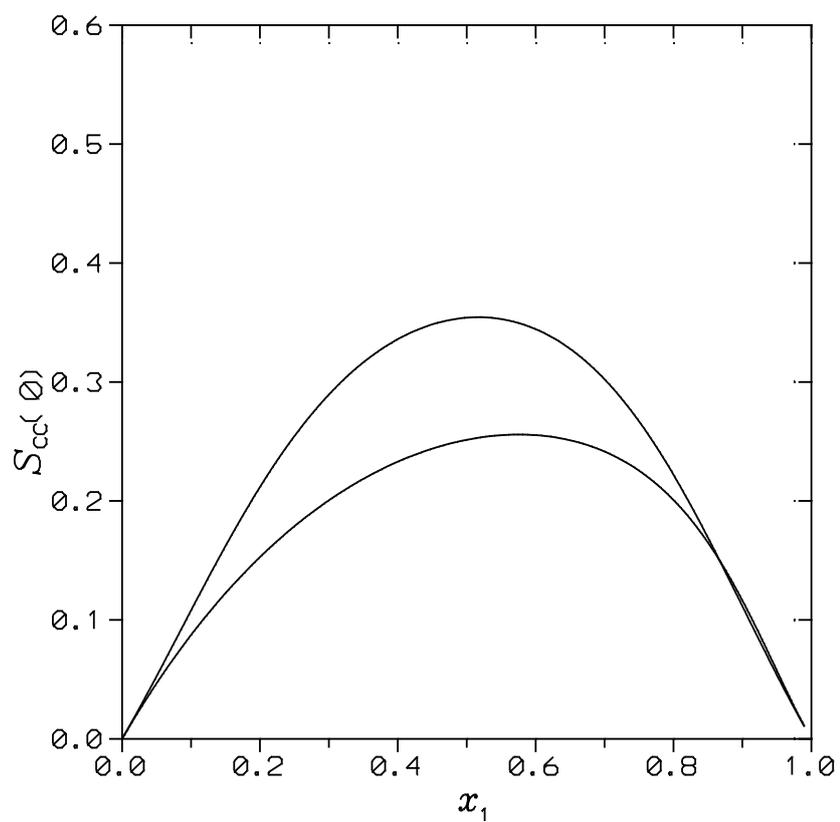

**Figure S2.** Experimental $S_{CC}(0)$ results for benzene(1) + $n$-alkane(2) mixtures at 298.15 K. Upper curve, heptane system; lower curve, tetradecane mixture